# Pseudo-Hermitian Systems Constructed by Transformation Optics with Robustly Balanced Loss and Gain

*Liyou Luo, Jie Luo\*, Hongchen Chu and Yun Lai\**

L. Luo, Dr. H. Chu, Prof. Y. Lai
National Laboratory of Solid State Microstructures
School of Physics
and Collaborative Innovation Center of Advanced Microstructures
Nanjing University
Nanjing 210093, China
E-mail: laiyun@nju.edu.cn

Dr. J. Luo
School of Physical Science and Technology
Soochow University
Suzhou 215006, China
E-mail: luojie@suda.edu.cn

Non-Hermitian systems with parity-time symmetry have been found to exhibit real spectra of eigenvalues, indicating a balance between the loss and gain. However, such a balance is not only dependent on the magnitude of loss and gain, but also easily broken due to external disturbance. Here, the authors propose a transformation-optics approach to construct a unique class of non-Hermitian systems with robustly balanced loss and gain, irrespective of the magnitude of loss/gain and the environmental disturbance. Through transformation-optics operators like space folding and stretching, loss and gain can be generated and separated in the real space. While in the virtual space, the loss and gain are still combined to each other, rendering a balance of energy that is far more robust than other non-Hermitian systems. This amazing feature is verified by finite-element simulations. This work reveals a class of non-Hermitian systems in which loss and gain are balanced robustly, thereby denoted as pseudo-Hermitian systems.

# 1. Introduction

Conservation of energy is a fundamental concept that shapes our understanding of physical reality. Canonical quantum mechanics focuses on closed systems where the Hamiltonian is Hermitian with real eigenvalues and conserved energy. However, in a limited subspace, energy or particles can exchange with its environment, and this subspace can be regarded as a non-Hermitian system. In general, non-Hermitian Hamiltonian possesses complex eigenvalues, implying non-conservation of energy. In 1998, Bender et al. [1] proved that a class of non-Hermitian Hamiltonians could exhibit real spectra if they commute with the parity-time (PT) operator, which require a complex potential $V(-\mathbf{r}) = V^*(\mathbf{r})$. In optics, the analog of the PT symmetry can be implemented by distributing gain and loss in space such that the refractive index profile satisfies $n(-\mathbf{r}) = n^*(\mathbf{r})$ [2-8]. A tremendous amount of interest is aroused in optical non-Hermitian systems, leading to many intriguing phenomena, such as optical exceptional points [9-17], coherent perfect absorption and lasing [18-22], advanced control of lasing [23-28], power oscillations [29-32], unidirectional invisibility [33-37], impurity-immunity effect [38], etc. However, although real eigenvalues are possible in non-Hermitian systems like PT symmetric ones, the conditions under which they appear are very strict. For instance, in PT-symmetric systems, when the magnitude of loss and gain exceeds some threshold, the system would transit from the PT exact phase to the PT broken phase, resurrecting the imbalance of loss and gain [9]. Moreover, in open systems, the PT symmetry could be easily broken by external disturbances, rendering the complete disappearance of PT exact phase with balanced loss and gain.

In this work, inspired by the concept of transformation optics (TO) [39, 40], we propose to construct a unique class of non-Hermitian optical systems in which loss and gain are robustly balanced. The idea comes from the analogy with the quantum field theory of vacuum, where virtual particles are created and annihilated, but not living long enough to be observed directly [41]. Imagine that if a system exhibits equal loss and gain at every point in space, then the electromagnetic (EM) energy would be generated and simultaneously absorbed at the same point, resulting in neither net loss nor net gain. Therefore, such a system appears Hermitian. Usually, if the loss and gain are separated and distributed differently in space, then the balance between loss and gain would be immediately broken and the system would become non-Hermitian. Intriguingly, space operators such as space folding and stretching in the TO provide a unique way to separate the loss and gain far apart in the physical space, yet still letting them overlap in the virtual optical space. Similar mechanisms have played important roles in many TO applications like invisibility cloaks [39, 42], cloaking at distance [43], illusion optics [44-46], etc. Unlike PT-symmetric systems, the loss and gain in the non-Hermitian systems generated by TO are always robustly balanced, not only irrespective of the magnitude of loss and gain, but also almost not influenced by external disturbance from the environment. Although the energy is generated, transported and absorbed at different locations in space, the amounts of generated and absorbed energy are always balanced internally within the system. This amazing feature of robust energy conservation bears a striking resemblance with the Hermitian system and inspires us to propose the concept of pseudo-Hermitian systems.

## 2. Pseudo-Hermitian Systems Constructed by TO Operators

In particle physics, any set of particles may be created and then annihilated. Interestingly, the total quantum numbers are zero as long as the conservation of energy and momentum are obeyed [41]. Here, we consider the EM analogy of this process. Figure 1(a) demonstrates the schematic graph of a special medium with equal amount of loss and gain everywhere in space. The electric field of an EM wave would induce the same amount of EM energy generation (red curved arrows) and absorption (blue curved arrows) simultaneously at every location. Therefore, the loss and gain of such a medium always cancel each other, giving a pure Hermitian response. If the gain and loss components of such a special medium are separated in space, then a gain medium and a lossy medium will be created simultaneously. Traditionally, this would change the original Hermitian system into a non-Hermitian one, because the generated energy in the gain medium is no longer required to be strictly balanced with the absorbed energy in the lossy medium. Only on very special conditions, such as the formation of the PT-symmetry with a limited magnitude of loss/gain, could the energy balance be reestablished again.

Interestingly, the theory of TO provides a unique approach to create a virtual optical space that appears to be completely different from the real space. For instance, TO has been demonstrated to change the optical appearance of one object in space and transform it into an almost arbitrary optical illusion [44-46]. With the help of TO, it is possible to create a special non-Hermitian system in which the gain and loss are separated in the real space, but still overlapping with each other in the virtual space. Therefore, the loss and gain can be robustly balanced as if they are still combined to each other. We denote such unique non-Hermitian systems generated by TO with robust balance of loss and gain as pseudo-Hermitian systems (Figure 1(a)). For instance, via the space folding operation [43], loss and gain overlapping in the virtual space can be separated in the real space. The concept is shown in Figure 1(b). Via the space stretching operation, adjacent loss and gain at the micro scale in the virtual space can also be pulled far away from each other in the real space. We emphasize that the robust energy balance in pseudo-Hermitian systems has no requirement on the limited magnitude of loss/gain as in PT-symmetric systems. Moreover, it is also not influenced by external disturbances from the environment. Such features are exceptional in all known non-Hermitian systems.

According to the TO theory, the permittivities of the media before and after the folding operation, i.e. $\varepsilon(\mathbf{x})$ and $\varepsilon(\mathbf{x}')$, are connected through $\varepsilon(\mathbf{x}') = \Lambda \varepsilon(\mathbf{x}) \Lambda^{\mathrm{T}} / \det \Lambda$, where $\Lambda$ is Jacobian transformation matrix $\Lambda_{ij} = \partial x_i' / \partial x_j$, and $\det \Lambda$ is the determinant of matrix $\Lambda$. The relationship between the permeabilities of the two media is similar, i.e. $\mu(\mathbf{x}') = \Lambda \mu(\mathbf{x}) \Lambda^{\mathrm{T}} / \det \Lambda$. The folding operation can be described by a mathematical mapping of $x' = -x$, $y' = y$, $z' = z$. Therefore, the anti-symmetric configuration of parameters is obtained as $\varepsilon(\mathbf{x}') = -\varepsilon(\mathbf{x})$ and $\mu(\mathbf{x}') = -\mu(\mathbf{x})$. We note that this relation is valid even when the parameters are complex numbers. This indicates that the lossy medium is transformed into the gain medium under the folding operation, and vice versa. On the other hand, the electric and magnetic fields in the two media before and after the transformation are connected by $\mathbf{E}(\mathbf{x}') = (\Lambda^{\mathrm{T}})^{-1} \mathbf{E}(\mathbf{x})$ and $\mathbf{H}(\mathbf{x}') = (\Lambda^{\mathrm{T}})^{-1} \mathbf{H}(\mathbf{x})$. Under the space folding operation, we obtain $E_{y(z)}(\mathbf{x}') = E_{y(z)}(\mathbf{x})$, $E_x(\mathbf{x}') = -E_x(\mathbf{x})$, and $H_{y(z)}(\mathbf{x}') = H_{y(z)}(\mathbf{x})$, $H_x(\mathbf{x}') = -H_x(\mathbf{x})$, indicating a mirror symmetric distribution with respect to the folding axis. Therefore, the magnitude of the EM fields on each point in the folded space $\mathbf{x}'$ is mathematically equal to those on the corresponding point in the original space $\mathbf{x}$.

Based on Poynting's theorem, the power dissipated at any point can be evaluated as,

$$-\text{Re}(\nabla \cdot \mathbf{S}) = \frac{1}{2}\omega\varepsilon_0\varepsilon''|\mathbf{E}|^2 + \frac{1}{2}\omega\mu_0\mu''|\mathbf{H}|^2 \qquad (1)$$

where $\mathbf{S} = (\mathbf{E} \times \mathbf{H}^*)/2$ is the complex Poynting vector. $\mathbf{E}$ ($\mathbf{H}$) is the electric (magnetic) field at this point. $\omega$ is the angular frequency. $\varepsilon''$ ($\mu''$) is the imaginary part of the relative permittivity (permeability). Considering the anti-symmetric $\varepsilon$ and $\mu$, together with symmetric $|\mathbf{E}|$ and $|\mathbf{H}|$ in the lossy and gain media, it is immediately obtained that EM energy generated at any point in the gain medium will be exactly absorbed (or compensated) at the corresponding point in the lossy medium. As a consequence, the total EM energy of the whole non-Hermitian system is robustly conserved, similar to the Hermitian systems without any gain or loss. Nevertheless, microscopically, there is a substantial amount of energy generated at the gain part, and then flow to the lossy part and be totally absorbed there. Therefore, the pseudo-Hermitian systems are not truly Hermitian.

## 3. Robust Energy Balance in Both Closed and Open Environments

In the following, we perform simulations by using the finite-element software COMSOL Multiphysics to demonstrate the robust EM energy balance in the proposed pseudo-Hermitian systems in both closed and open environments. For comparison, we have also demonstrated the energy imbalance in corresponding PT-symmetric systems under the same disturbance.

Firstly, we investigate the circular pseudo-Hermitian model consisting of a semicircular gain medium (left) and a semicircular lossy medium (right) enclosed by a square of perfect electric conductor (PEC) with a side length of $a$. The schematic graph of this closed system is illustrated in Figure 2(a). The diameter of the circular pseudo-Hermitian model is $0.4a$. The relative permittivity $\varepsilon_g$ of the gain medium is complex (i.e. $\varepsilon_g = \varepsilon' - i\varepsilon''$ with $\varepsilon'$ and $\varepsilon''$ being the real and imaginary parts, respectively). The relative permittivity $\varepsilon_l$ of the lossy medium is thus $\varepsilon_l = -\varepsilon_g = -\varepsilon' + i\varepsilon''$. Here, we fix $\varepsilon' = 2.25$ and $\mu = 1$, while vary the $\varepsilon''$ from 0 to 5. The relative permeabilities of the lossy and gain media are set as real numbers $\mu_g = -\mu_l = 1$. In this way, the gain medium represents a normal dielectric media with gain, and the lossy medium is a left-handed medium [47, 48] with loss.

In Figure 2(b), we present the normalized eigen-frequency $fa/c$ of the pseudo-Hermitian system in a closed environment. Here, $f$ is the eigen-frequency, and $c$ is the speed of light in vacuum. The eigen-frequencies of the first two eigenmodes of transverse-electric (TE)-polarization (with out-of-plane electric fields) are plotted as solid/dashed lines with solid/hollow dots. The two modes are distinguished by different colors. From Figure 2(b), it is clearly seen that the eigen-frequencies remain as real values regardless of the variation of $\varepsilon''$, indicating that the energy is always conserved in such a non-Hermitian system.

For comparison, in Figure 2(c), we replace the pseudo-Hermitian system with a PT-symmetric one. The parameters of the gain medium are not changed, while those of the lossy medium are changed into $\varepsilon_l = \varepsilon_g^* = \varepsilon' + i\varepsilon''$ and $\mu_l = \mu_g^* = 1$, which indicate a normal dielectric medium with loss. The normalized eigen-frequencies of the first two TE modes are plotted in Figure 2(d). Clearly, there is phase transition behavior at the exceptional point [4] at $\varepsilon'' = 2.75$. Below the exceptional point, the eigen-frequencies are simply real, indicating the system is in the PT exact phase (green region) where loss and gain are always balanced. However, when $\varepsilon''$ is above the exceptional point, the eigen-frequencies become complex (gray region) and the system lies in the PT-broken phase. The modes turn into two asymmetric ones which are dominated by energy amplification and absorption, respectively. The energy balance realized in PT-symmetric systems is clearly not as robust as that in pseudo-Hermitian ones.

Next, we investigate the influence of environment disturbances in this closed system. The gain/lossy medium pair in Figure 2(a) and 2(c) is moved to the left over a distance of $0.1a$. We note that despite that such a change is tiny, the spatial symmetry of the whole system is broken. Figure 2(e) and 2(g) show schematic graphs of the pseudo-Hermitian model and PT-symmetric models, respectively. The corresponding normalized eigen-frequencies of the two systems are plotted as the function of $\varepsilon''$ in Figure 2(f) and 2(h), respectively. It is clearly seen that the EM energy is still well conserved in the pseudo-Hermitian model, as implied by the pure real eigen-frequencies. While the eigen-frequencies of the PT-symmetric model change dramatically and the PT exact phase with real spectra completely vanishes, as a consequence of the broken PT symmetry for the whole system. These results show that the conservation of energy in the pseudo-Hermitian system is very robust to the environmental disturbances, because the virtual space generated by the TO theory is not influenced by environments. In contrast, the PT-symmetric system is very sensitive to the environment as long as the fields are not locally confined as if in waveguide systems [9-13]. Thus, the environmental disturbance can break the PT symmetry and lead to the amplification or decay of EM energy.

Next, we investigate the energy balance properties of the pseudo-Hermitian and PT-symmetric models under the illumination by a TE-polarized plane wave in an open environment, as shown in Figure 3. The operating frequency is chosen as $8c/15a$, and the incident angle is $\varphi$. In the simulations, perfectly matched layers (PMLs) are applied to avoid reflection from the boundaries of the simulation area. To evaluate the EM energy in this open system, we calculate the power generated within the gain media $P_g$, and the power absorbed within the lossy medium $P_l$ as,

$$P_g = \frac{1}{2}\int_g \omega\varepsilon_0\varepsilon''|\mathbf{E}_g|^2 \mathrm{d}S, \text{ and } P_l = \frac{1}{2}\int_l \omega\varepsilon_0\varepsilon''|\mathbf{E}_l|^2 \mathrm{d}S, \qquad (2)$$

where $\mathbf{E}_g$ and $\mathbf{E}_l$ are electric fields in the gain and lossy media, respectively. Figure 3(b) presents the power ratio $P_l/P_g$ of the pseudo-Hermitian model as a function of incident angle $\varphi$ when $\varepsilon''$ is set as 0.1, 0.2, 0.5, 1, 2 and 5, respectively. It is seen that this power ratio is always near unity for different values of $\varphi$ and $\varepsilon''$, and the electric fields are symmetrically distributed in the gain and lossy media (the inset graph). This indicates that the EM energy generated by the gain medium is always absorbed by the lossy medium even in open environments. In contrast, the power ratio $P_l/P_g$ in the PT-symmetric system illustrated in Figure 3(c) is generally different from unity, and varies significantly with the change of $\varphi$ and $\varepsilon''$, as seen in Figure 3(d). The electric fields are obviously asymmetric in the gain and lossy media (see the inset), thus the EM energy is imbalanced.

In order to further demonstrate the influence of environmental disturbances to the energy balance in this open system, we add a square dielectric scatterer (relative permittivity 10, side length $0.1a$) on the top left side beside the loss/gain media, as illustrated in Figure 3(e) for the pseudo-Hermitian model and Figure 3(g) for the PT-symmetric model, respectively. The power ratio $P_l/P_g$ corresponding to the two systems are plotted in Figure 3(f) and 3(h), respectively. Despite of the existence of the dielectric scatterer, the power ratio in the pseudo-Hermitian model is robustly near unity, as shown in Figure 3(f). The electric fields are still constrained by the TO theory and remain symmetric in the gain and lossy media despite of the scatterer. On the other hand, the power ratio in the PT-symmetric model is dramatically changed by the dielectric scatterer. This is because the fields inside the loss/gain media are largely influence by the scatterer, as shown in Figure 3(h). The results again prove that the energy balance in the pseudo-Hermitian systems is much more robust that that in the PT-symmetric ones.

## 4. More Examples of TO Operators for Pseudo-Hermitian Systems

The space-folding transformation is not the only TO operator that can realize pseudo-Hermitian systems. Space stretching transformations can also separate two adjacent points far apart in the real space, while keeping them adjacent in the virtual space. In the following, we demonstrate pseudo-Hermitian systems constructed by the space stretching transformations.

In the example, we assume that in the virtual space the generation and absorption processes of EM energy are simultaneously happening on a circular thin film. The magnitudes of the two processes are exactly equal to each other. The separation distance between the two processes are set to be negligible, i.e. the thin gain film and absorptive film are nearly overlapping with each other. Due to the nearly overlapping loss and gain of the same magnitude, robust energy conservation as if in Hermitian systems can be achieved. The model of the system in the virtual space is shown in Figure 4(a). In order to prove the Hermitian-like optical response, we illuminate the system by TE-polarized Gaussian beams with a wavelength of $\lambda$ and a distance $d$ from the center of the system. The Rayleigh range of the beam is $20\lambda/3$. The inner radius of the gain film is $R = 300\lambda$, the thickness of the gain film is $d_g = \lambda/60$, and the gap between the gain and lossy films is set to be $d_{gap} = 10^{-4}\lambda$, thus the outer radius of the lossy film is $R_2 = R - d_{gap}$. The thickness of the lossy film, i.e. $d_l = R_2 - \sqrt{R_2^2 - 2Rd_g - d_g^2}$, is obtained by assuming the equality of the areas of gain and lossy films. In Figure 4(a), the sizes of the gain and lossy films are magnified for illustration of the model. We assume that both the gain and lossy films are nonmagnetic, and have relative permittivities $\varepsilon_g = 2.25 - i\varepsilon''$ and $\varepsilon_l = 2.25 + i\varepsilon''$, respectively. The power ratio $P_l/P_g$ of such a system in the virtual space is calculated and plotted in Figure 4(b), as a function of normalized incident position $d/R$ and $\varepsilon''$. Clearly, $P_l/P_g$ is almost unity, indicating robust balance between the gain and loss in the films.

In the next, we apply the TO operator to form a cylindrical concentrator [49], at the same time stretching the gain and lossy films far apart from each other. The space mapping is

$$r' = \begin{cases} \dfrac{R_1}{R_2}r, & 0 < r < R_2 \\ \dfrac{R - R_1}{R - R_2}(r - R_2) + R_1, & R_2 < r < R \end{cases}$$
$$\theta' = \theta \qquad (3)$$
$$z' = z$$

where $R_1 = 2R/5$. After the transformation, the separation distance between gain and lossy films is around $R - R_1 = 3R/5$, which is significantly larger than before. Two types of TO media are formed in the regions I and II, respectively. In the region I, the space is compressed, while in the region II, the space is stretched. The detailed parameters can be obtained from Equation (3). The parameters and shape of the lossy film are also changed accordingly under the rules of the TO compressing operator. Therefore, a unique pseudo-Hermitian system is constructed, which consists of a pair of circular gain and lossy thin films as well as both compressed and stretched TO media. By illuminating the Gaussian beam onto this pseudo-Hermitian system, we find that the optical response is almost exactly the same as the original system in the virtual space. The power ratio $P_l/P_g$ is almost unity, irrespective of the normalized incident position $d/R$ and $\varepsilon''$. The electric field maps are shown by the insets in Figure 4(b) and 4(d). The electric fields on the gain film are transferred to the lossy film by the TO media in the region II at the same magnitude and phase. This leads to the robust energy balance between the two films. In contrary, if we change the two TO media in the regions I and II to free space, as schematically shown in Figure 4(e), the balance between gain and lossy films will be immediately destructed. This is proven by the dramatically increased or reduced power ratio $P_l/P_g$, which now depends on $d/R$ and $\varepsilon''$, as shown in Figure 4(f).

## 5. Conclusions

In short, our work reveals that TO can generate a new class of non-Hermitian optical systems denoted as pseudo-Hermitian systems, which exhibit unprecedented robustness in energy balance, far superior to any other non-Hermitian systems, including PT-symmetric ones. The inspiration of our work comes from the establishment of a virtual optical space with balanced loss and gain at every location, which is an analog of the quantum field theory of vacuum, where virtual particles are created and annihilated, but not living long enough to be observed directly. The robust energy balance in pseudo-Hermitian systems may inspire novel applications such as the transport of electromagnetic wave energy with near 100% efficiency.

**Supporting Information**

Supporting Information is available from the Wiley Online Library or from the author.

**Acknowledgements**

The authors thank Prof. Jensen Li for helpful discussions. This work is supported by National Key R&D Program of China (2017YFA0303702), National Natural Science Foundation of China (11704271, 61671314, 11974176), Natural Science Foundation of Jiangsu Province (BK20170326), and a Project Funded by the Priority Academic Program Development of Jiangsu Higher Education Institutions (PAPD).

**Conflict of Interest**

The authors declare no conflict of interest.

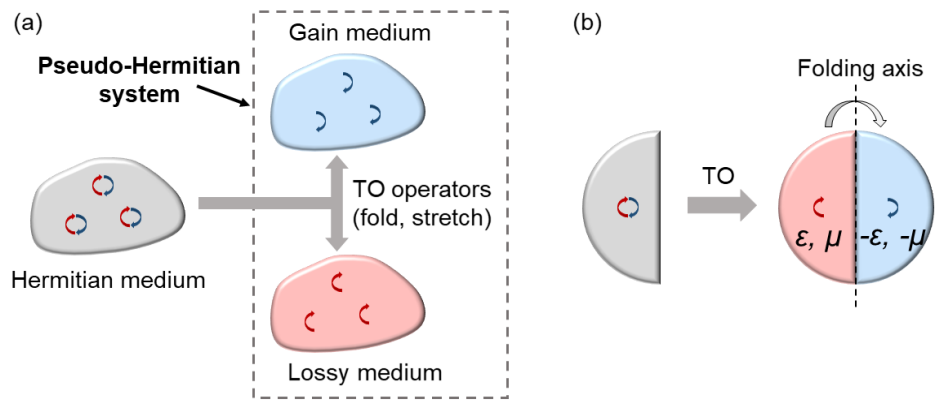

**Figure 1** (a) Illustration of the construction of energy-balanced pseudo-Hermitian systems by using TO operators to separate the energy generation and absorption processes in a Hermitian media. (b) Illustration of a pseudo-Hermitian system constructed by applying the space-folding transformation in TO.

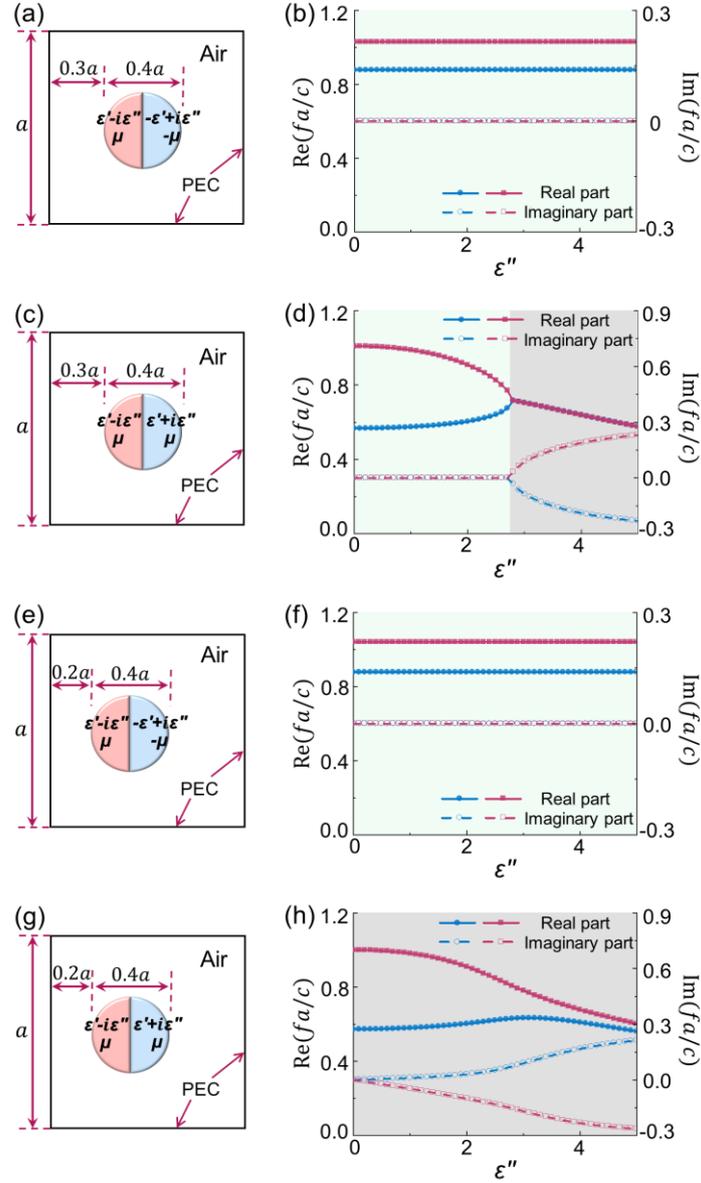

**Figure 2** Schematic graphs of the [(a) and (e)] pseudo-Hermitian and [(c) and (g)] PT-symmetric systems bounded by a square of PEC wall. The centers of the systems have a position offset of $0.1a$ to the left in (e) and (g). [(b), (d), (f) and (g)] Normalized eigen-frequencies of the first two TE modes as the function of $\varepsilon''$ in the cases of (a), (c), (e) and (g), respectively. The solid (dashed) lines with dots denote the real (imaginary) part of the eigen-frequencies. The two modes are distinguished by purple and blue colors. The green and gray regions denote the regimes of energy balance and imbalance, respectively.

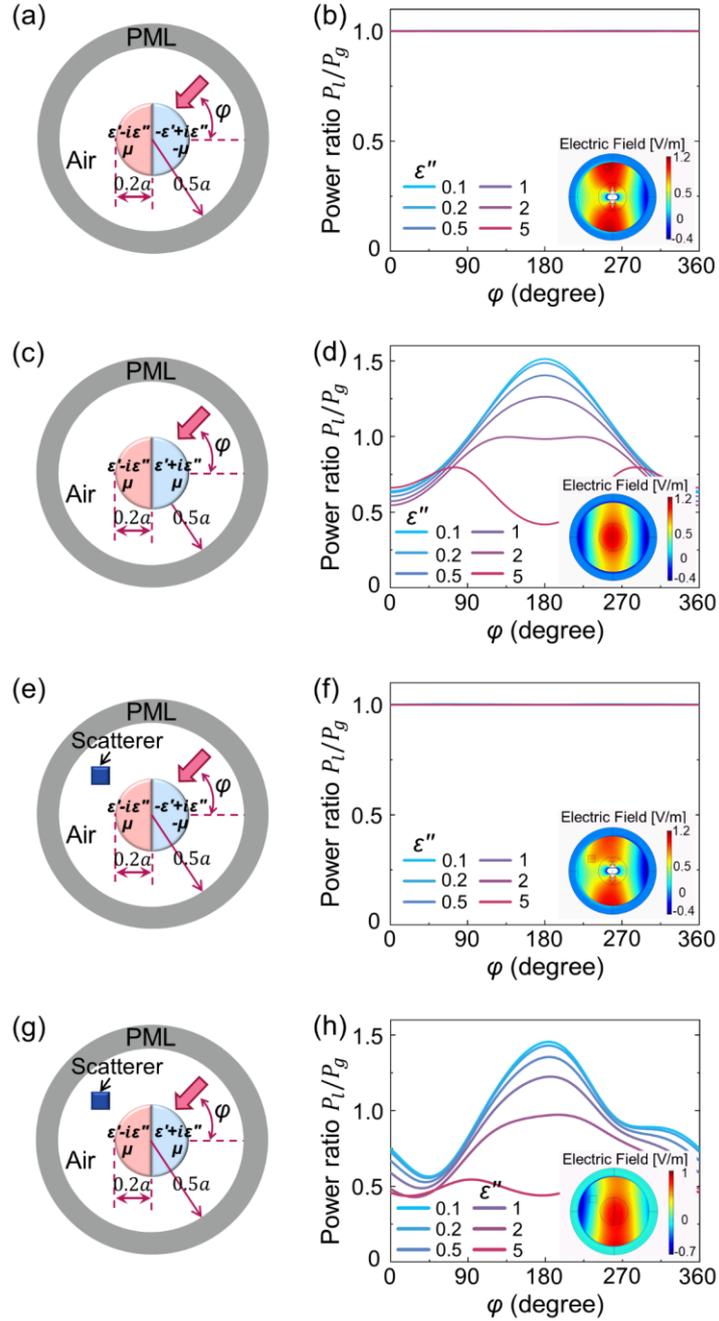

**Figure 3** Schematic graphs of [(a) and (e)] pseudo-Hermitian and [(c) and (g)] PT-symmetric systems in an open environment under the illumination by a TE-polarized plane wave with an incident angle of $\varphi$. A square dielectric scatterer is placed in the top left side in (e) and (g). [(b), (d), (f) and (g)] Power ratio $P_P/P_A$ as functions of the incident angle $\varphi$ and $\varepsilon''$ in the cases of (a), (c), (e) and (g), respectively. The insets show the electric-field distributions under $\varphi = 0$ and $\varepsilon'' = 0.5$.

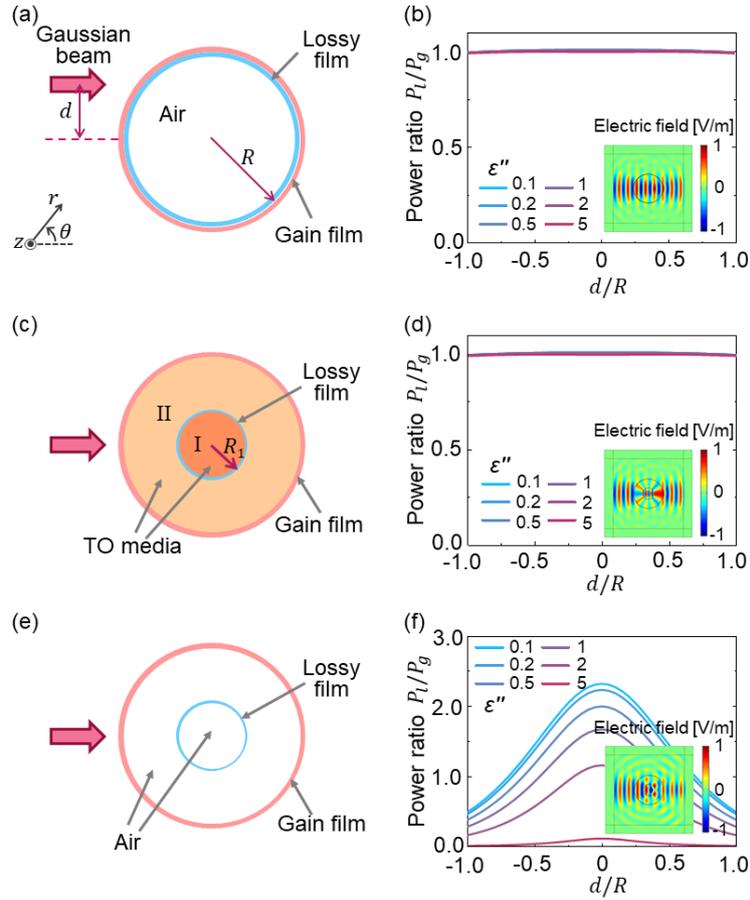

**Figure 4** [(a), (c) and (e)] Illustrations of (a) the virtual optical space, (c) pseudo-Hermitian system formed by the TO operator, and (e) non-Hermitian system with the TO media replaced by free space. [(b), (d) and (f)] Power ratio $P_l/P_g$ obtained under the illumination by a TE-polarized Gaussian wave as functions of $d/R$ and $\varepsilon''$ in the cases of (a), (c) and (e), respectively. The insets show the electric field-distributions under the conditions of $d/R = 0$ and $\varepsilon'' = 0.1$.